\documentstyle[12pt]{article}
\title{Stationary mKdV hierarchy and integrability of
the Dirac equations by quadratures}
\author{R.Z.~Zhdanov\thanks{e-mail: rzhdanov@apmat.freenet.kiev.ua},
\ I.V.Revenko and W.I.~Fushchych \\ \small Institute
of Mathematics of the Academy of Sciences of Ukraine,\\
\small Tereshchenkivska Street 3, 252004 Kiev, Ukraine}
\date{}
\newtheorem{theo}{Theorem}
\newtheorem{lem}{Lemma}
\begin{document}
\maketitle
\begin{abstract}
Using the Lie's infinitesimal method we establish that
the Dirac equation in one variable is integrable by
quadratures if the potential V(x) is a solution of
one of the equations of the stationary mKdV hierarchy.
\end{abstract}

Consider the eigenvalue problem for the Dirac operator ${\cal L} =
i\sigma_1 {d \over dx} - V(x) \sigma_2$
\begin{equation}
\label{dir}
({\cal L} - \lambda)\vec u\equiv i\sigma_1{d \vec u \over dx} -
(V(x) \sigma_2 + \lambda)\vec u = \vec 0
\end{equation}
where $\sigma_1, \sigma_2$ are $2\times 2$ Pauli matrices,\
$\vec u = (u_1(x),u_2(x))^{T}$, $V(x)$ is a real-valued function and $\lambda$
is a real parameter. We remind that (\ref{dir}) is one of two equations composing
the Lax pair for the mKdV equation
\begin{equation}
\label{mkdv}
v_t + v_{xxx} - 6v^2 v_{x}=0
\end{equation}
integrable by the inverse scattering method (see, e.g. \cite{zah,lamb}). Next, as
the identity
\[
({\cal L} - \lambda)({\cal L} + \lambda) =-{d^2\over dx^2} + V^2 - \sigma_3
{d V\over dx} - \lambda^2
\]
holds, components of the vector-function $\vec u$ fulfill the stationary
Schr\"odinger equation
\begin{equation}
\label{sch}
{d^2 u_i\over dx^2} + \left((-1)^{i+1}{dV\over dx} - V^2
+ \lambda^2\right)u_i=0,\quad i=1,2.
\end{equation}

The aim of the present paper is to show that there exists an initimate
connection between integrability of system (\ref{dir}) (in what follows we will
call it the Dirac equation) by quadratures and solutions of the
stationary mKdV hierarchy.

Integrability of system (\ref{dir}) will be studied with the use of
its Lie symmetries. As usual, we call a first-order differential operator
\[
X=\xi(x){d \over dx} + \eta(x),
\]
where $\xi$ is a real-valued function and $\eta$ is a $2\times 2$ matrix
complex-valued function, a Lie symmetry of system (\ref{dir}) if commutation
relation
\begin{equation}
\label{com}
[{\cal L},\ X]=R(x){\cal L}
\end{equation}
holds with some $2\times 2$ matrix function $R(x)$ (for details, see, e.g. \cite{fun}).

A simple computation shows that if $X$ is a Lie symmetry of system (\ref{dir}),
then an operator $X + r(x){\cal L}$ with a smooth function $r(x)$ is its
Lie symmetry as well. Hence we conclude that without loss of generality
we can look for Lie symmetries within the class of matrix operators $X=\eta(x)$.
Furthermore, inserting $X=\eta(x)$ into (\ref{com}) and computing the commutator
yield that the matrix $\eta(x)$ is necessarily of the form
\begin{equation}
\label{eta}
\eta=\left\|\begin{array}{cc} f(x)& g(x)\\
                              h(x)& -f(x)\end{array}\right\|,
\end{equation}
where $f(x), g(x), h(x)$ are arbitrary solutions of the following
system of ordinary differential equations:
\begin{equation}
\label{det}
{df\over dx}=i\lambda (g-h),\quad {dg\over dx}= 2i\lambda f + 2gV,\quad
{dh\over dx} = -2i\lambda f - 2hV.
\end{equation}

With a solution of system (\ref{det}) in hand we can integrate
the initial equations (\ref{dir}) by quadratures using the classical
results by Elie Cartan \cite{car}. Since these results are well-known
we will give them without derivation in the form of the following lemma.
\begin{lem}
Let the functions $f(x), g(x), h(x)$ satisfy system (\ref{det}). Then the
general solution of the Dirac equation is given by the formulae
\begin{equation}
\label{sol}
\begin{array}{rcl}
u_1(x)&=&C_1\, (R(x) + f(x))\,(h(x))^{-1/2}\, (R^2(x) - \Delta)^{-1/2},\\
u_2(x)&=&C_1\, (h(x))^{1/2}\, (R^2(x) - \Delta)^{-1/2},
\end{array}
\end{equation}
where $\Delta = f^2(x)+ g(x)h(x)$ is constant on the solution variety of
system (\ref{det}),
\[
R(x)=\left\{\begin{array}{ll} \sqrt{\Delta}\tanh\left(C_2 - i\lambda \sqrt{\Delta}
\displaystyle\int {\displaystyle dx\over \displaystyle g(x)}\right),&{\rm under}\
\Delta > 0,\\
\left(C_2 - i\lambda
\displaystyle\int {\displaystyle dx\over \displaystyle g(x)}\right)^{-1},&{\rm under}\
\Delta = 0,\\
\sqrt{-\Delta}\tan\left(C_2 + i\lambda \sqrt{-\Delta}
\displaystyle\int {\displaystyle dx\over \displaystyle g(x)}\right),&{\rm under}\
\Delta < 0
\end{array}\right.
\]
and $C_1, C_2$ are arbitrary complex constants.
\end{lem}

However, solving system of ordinary differential equations (\ref{det})
is by no means easier than solving the initial Dirac equation. This is
a common problem in applying Lie symmetries to integration of ordinary
differential equations. The key idea of our approach is to restrict
{\em a priori}\ the class within which Lie symmetries are looked for
and suppose that they are polynomials in $\lambda$ with
variable matrix coefficients.

Introducing the new dependent variables $\psi_1(x), \psi_2(x)$
\begin{eqnarray}
f(x)&=&\frac{i}{4\lambda}\left(-{d\psi_1\over dx} + 2V\psi_2\right),\nonumber\\
g(x)&=&\frac{1}{2}(\psi_1(x) + \psi_2(x)),\label{sym}\\
h(x)&=&\frac{1}{2}(\psi_1(x) - \psi_2(x))\nonumber
\end{eqnarray}
we rewrite (\ref{det}) in the following equivalent form:
\begin{equation}
\label{det1}
{d^2\psi_1\over dx^2} = -4\lambda^2 \psi_1 + 2V {d\psi_2\over dx}
+ 2\psi_2 {dV\over dx},\quad {d\psi_2\over dx}= 2 V \psi_1.
\end{equation}

As mentioned above solutions of system (\ref{det1}) are looked for
as polynomials in $\lambda$, namely
\begin{equation}
\label{an}
\psi_1(x)=\sum_{k=1}^{n}p_k(x)(2\lambda)^{2k},\quad
\psi_2(x)=\sum_{k=1}^{n}r_k(x)(2\lambda)^{2k}.
\end{equation}

Inserting the expressions (\ref{an}) into (\ref{det1}) and equating the coefficients
by the powers of $\lambda$ yield $p_n=0$ and
\begin{eqnarray}
{dr_k\over dx} &=& 2 V p_k,\quad k=1,\ldots,n\label{eq1}\\
{d^2p_k\over dx^2} &=& 2 V {dr_k\over dx} + 2 {dV\over dx}r_k - p_{k-1},\quad
k=1,\ldots,n-1,\label{eq2}
\end{eqnarray}
where we set by definition $p_{-1}(x)=0$. Eliminating from (\ref{eq1}),
(\ref{eq2}) the functions $r_k(x)$ we get recurrent relations for the
functions $p_k(x)$:
\begin{equation}
\label{rec}
p_{k-1}(x)=\underbrace{\left\{-{d^2\over dx^2} +
4{d V\over dx} D_x^{-1} V + 4V^2\right\}}_{\cal Q}p_k(x),\quad k=n,n-1,\ldots,0.
\end{equation}
Here $D_x^{-1}$ denotes integration by $x$.

A reader familiar with the theory of solitons will immediately
recognize the operator ${\cal Q}$ as the recursion operator for
the mKdV equation (\ref{mkdv}) (see, e.g. \cite{ibr,olv}). Acting repeatedly with this
operator on the trivial symmetry $S_0=0$ yields
an infinite number of higher symmetries $S_1, S_2,\ldots$ admitted by
the mKdV equation \cite{ibr}. Hence it is not difficult to derive that
the functions $p_k, k=0,\ldots,n-1$ are linear combinations
of the higher symmetries $S_1,\ldots,S_{n}$ with arbitrary
constant coefficients $C_i$
\begin{equation}
\label{coe}
p_{n-k}(x)=\sum_{i=1}^kC_{i}S_{k+1-i},\quad k=1,\ldots,n,
\end{equation}
where $S_i$ are determined by the recurrent relations
\[
S_{i+1}(x)=-{d^2S_i(x)\over dx^2} + 4{d V\over dx} \int\limits_{x_0}^xV(y)S_i(y)dy
+ 4V^2S_i(x),\quad i=1,\ldots,n-1
\]
with $S_1\stackrel{{\rm def}}{=}{dV\over dx}$.

The above formulae (\ref{coe}) give the general solution of the first $n$
equations from (\ref{rec}). Inserting these into the last equation yields
equation for the function $V(x)$ of the form
\begin{equation}
\label{hie1}
\sum_{k=1}^{n+1}C_{k}S_{n+2-k} = 0.
\end{equation}
As $S_1={dV\over dx}$, (\ref{hie1}) is nothing else than an equation
of the stationary mKdV hierarchy, which is obtained from the higher
mKdV equations by setting $v(t,x)=v(x+Ct),\ C={\rm const}$.

Integrating equations (\ref{eq1}) yields
\begin{equation}
\label{coe2}
r_k(x)= 2\sum_{i=1}^kC_{i}\int\limits_{x_0}^xV(y)S_{k+1-i}(y)dy + \tilde C_k,
\quad k=1,\ldots,n,
\end{equation}
where $\tilde C_i$ are arbitrary complex constants.

Thus, the formulae (\ref{an}), (\ref{coe}), (\ref{hie1}), (\ref{coe2}) give the general
solution of the system of determining equations (\ref{eq1}), (\ref{eq2})
within the class of functions of the form (\ref{an}). This means, in particular,
that provided the function $V(x)$ is a solution of equation (\ref{hie1})
with some fixed $n$ and $C_1,\ldots,C_n$, the Dirac equation possesses
a Lie symmetry. Hence we conclude that it is integrable by quadratures
due to Lemma 1. Consequently, we have proved the following remarkable fact.
\begin{theo}
Let $V(x)$ be a solution of an equation of the mKdV hierarchy of the
form (\ref{hie1}). Then the Dirac equation (\ref{dir}) is integrable
by quadratures.
\end{theo}

Note that the equations of the stationary mKdV hierarchy are
transformed to the equations of the stationary KdV hierarchy with the
help of the Miura transformation and the latter are integrated in
$\theta$-functions with any $n\in {\bf N}$ \cite{nov}.

In conclusion let us demonstrate how the above procedure works for the simplest
case $n=1$. With this choice of $n$ equation (\ref{hie1}) reads
\begin{equation}
\label{smkdv}
\frac{C_2}{C_1}{dV\over dx} - {d^3V\over dx^3} + 6 V^2{dV\over dx} = 0
\end{equation}
which is exactly the stationary mKdV equation and is obtained from (\ref{mkdv})
via the Ansatz $v(t,x)=V(C_2x - C_1t)$.

A simple computation yield the form of the coefficients of
the Lie symmetry (\ref{eta})
\begin{eqnarray}
f(x)&=&-\frac{i}{4\lambda}\left(C_1 {d^2 V\over dx^2} - 2C_1V^3 - C_2
-4C_1\lambda^2)\right),\nonumber\\
g(x)&=&\frac{1}{2}\left(C_1{dV\over dx} - C_1V^2 - \frac{1}{2}C_2
- 2C_1\lambda^2\right),
\label{last}\\
h(x)&=&\frac{1}{2}\left(C_1{dV\over dx} + C_1V^2 + \frac{1}{2}C_2
+ 2C_1\lambda^2\right)
\nonumber
\end{eqnarray}
which satisfy the determining equations (\ref{det}) inasmuch as the function
$V(x)$ is a solution of the stationary mKdV equation.

Thus, the Dirac equation with potential $V(x)$ satisfying the stationary
mKdV equation (\ref{smkdv}) is integrable by quadratures and its general
solution is given by formulae (\ref{sol}), (\ref{last}).

Note that due to the remark made at the very beginning of the
paper components of the function $\vec u$ fulfill the stationary
Schr\"odinger equation (\ref{sch}). This is in a good accordance with results
of papers \cite{zhd,doe}.

\end{document}